\documentclass[%
reprint,
superscriptaddress,
 amsmath,amssymb,
 aps,
prb,
twocolumn
]{revtex4-2}

\usepackage{graphicx,bm,braket,color,hyperref,comment}
\usepackage{fancybox}
\usepackage{ulem}
\hypersetup{colorlinks=true,citecolor=blue,linkcolor=blue,urlcolor=blue}
\begin{document}


\title{Superzone Gap Formation Induced by Ferroic Orders
}

\author{Takayuki Ishitobi}
\affiliation{Advanced Science Research Center, Japan Atomic Energy Agency, Tokai, Ibaraki 319-1195, Japan}

\date{\today}

\begin{abstract}
We demonstrate that a superzone gap, typically associated with antiferroic ordering, can also emerge from ferroic orders in systems with sublattice degrees of freedom. By analyzing a $p$-orbital tight-binding model on a zigzag chain, we show that a Su--Schrieffer--Heeger-type gap is induced by ferroquadrupole or ferromagnetic order or by applying an external magnetic field. 
\end{abstract}

\pacs{Valid PACS appear here}
\maketitle


Understanding the spontaneous symmetry breaking associated with ordering and its impact on physical properties is a fundamental issue in condensed matter physics. A representative example is the formation of a so-called superzone gap, where parts or all of the Fermi surface vanish due to Peierls transitions or antiferromagnetic ordering, among others \cite{Ashcroft_Mermin}. This leads to macroscopic phenomena, such as an increase in resistivity below the transition temperature. 
Electrical resistivity, one of the most basic and accessible measurements after sample synthesis, can thus provide critical insights into the nature of the ordered phases. This is particularly useful for elucidating ordering mechanisms and exploring new materials. 

However, in particular compounds such as ${\rm UCu_2Sn}$ and CeCoSi, the resistivity behavior indicative of superzone gap formation led to their initial identification as antiferromagnetic or antiferroquadrupolar orders~\cite{Takabatake1998-ol, Lengyel2013-tn}, which were later identified as ferroquadrupole orders~\cite{Suzuki2000-kb, Matsumura2022-ky}. The origin of this contradiction remains unclear. 
This discrepancy presents a fundamental challenge to the scheme of inferring order parameters from accessible physical measurements. In this study, we investigate the formation of band gaps at the Brillouin zone boundary driven by ferroic ordering. 

\begin{figure}[t!]
\centering
\includegraphics[width=0.45\textwidth]{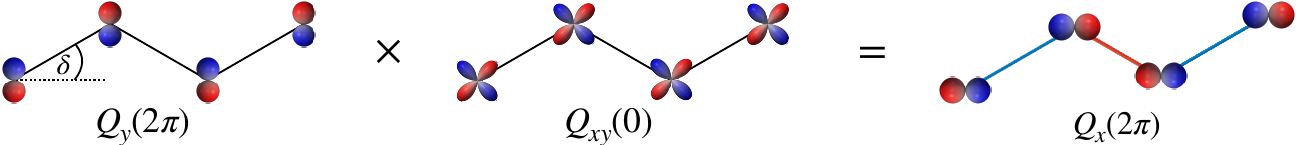}
\caption{(Color online) Schematic illustration of bond-order induction due to a ferroquadrupole order on a zigzag chain. Red and blue indicate the sign of charge or bond modulation.}
\label{fig:Schematics}
\end{figure}

Let us first review the mechanism of gap formation and degeneracy lifting. As discussed in standard textbooks, the static potential at a reciprocal lattice vector ${\bm G}$ component opens a gap at Brillouin zone boundary ${\bm k} = {\bm G}/2$~\cite{Ashcroft_Mermin}. 
In systems with sublattice degrees of freedom, for example, in a spinless single-orbital model on a one-dimensional zigzag chain, a degeneracy exists at the zone boundary $k=\pi$. The zigzag chain is characterized by alternating atomic displacements in the direction ($y$) perpendicular to the chain ($x$), corresponding to an antiferroic ordering of electric dipoles $Q_y(2\pi)$, as shown in Fig.~\ref{fig:Schematics} left. Since electric dipoles do not couple to charge—the only degree of freedom in a spinless single-orbital model—no hybridization occurs at $k = \pm \pi$. 
In contrast, the dipoles alternate along the chain, i.e., ordering of $Q_x(2\pi)$, corresponds to bond alternation, as shown in Fig.~\ref{fig:Schematics} right, and results in a gap opening as in the Su-Schrieffer-Heeger model~\cite{Su1979-ks}. This can be understood because longitudinal atomic displacements lead to modulation of charge density at the bond centers and couple to the system's degrees of freedom. 

In general, degeneracy at the zone boundary ${\bm k}_{1,2}$, typically ${\bm k}_2 = -{\bm k}_1$, due to sublattice degrees of freedom can be lifted when a charge or a longitudinal electric dipole orders with wavevector ${\bm q}={\bm k}_1-{\bm k}_2$. If orbital and/or spin degrees of freedom are present, the off-diagonal sector of ${\bm k}_{1,2}$ becomes matrix-valued and can be expressed in a multipole basis~\cite{Hayami2024-vw}. In such cases, the degeneracy is partially or fully lifted when a multipole moment in the degenerating sector, or product of it with a longitudinal dipole, orders at ${\bm q}$. In terms of periodicity, a superzone gap arises when the periodicity of the degenerating multipole sector, rather than all sectors, is extended. 

Considering this general framework, we examine gap formation at the zone boundary induced by ferroic orders. As the simplest system with sublattice degrees of freedom, we analyze a one-dimensional zigzag chain as a minimal model. The zigzag chain exhibits antiferroic ordering of transverse dipoles $Q_y(2\pi)$. When a ferroic quadrupole $Q_{xy}(0)$ orders, the product of this quadrupole with the transverse dipole $Q_y(2\pi)$ yields a longitudinal dipole $Q_x(2\pi)$, thereby inducing a bond-alternating order (Fig.~\ref{fig:Schematics}) and opening a band gap. We demonstrate this mechanism explicitly and also show that ferromagnetic order or $xy$-plane magnetic fields can cause the gap opening, as they induce $Q_{xy}(0)$ at their second order. 

To represent quadrupolar degrees of freedom, we adopt a tight-binding model of $p$ orbitals ($p_x$, $p_y$, $p_z$) on a zigzag chain. The Hamiltonian is given by
\begin{align}
H = \sum_{\langle i,j \rangle} H_{ij}^{\rm hop} 
+ \sum_i ( \lambda{\bm L} \cdot {\bm \sigma} - \Delta Q_{xy} - {\bm B}\cdot {\bm M})
\label{eq:H_r}
,\end{align}
where $i,j$ label sites on the zigzag chain. The first term is the nearest-neighbor $\sigma$-bond hopping, the second is the spin-orbit coupling, the third represents mean-field interaction from quadrupole order, and the fourth is the Zeeman term. The quadrupole moment $Q_{xy} = -(L_xL_y + L_yL_x)$ and magnetic moment ${\bm M} = {\bm L} + {\bm \sigma}$ are defined in terms of the orbital and spin angular momentum operators ${\bm L}$ and ${\bm \sigma}/2$. 

The hopping $t_{mn}$ between $p_m$ and $p_n$ orbitals for the nearest-neighbor bond of angle $\pi - 2\delta$ are given by 
$t_{xx} = t\cos^2\delta$, $t_{yy} = -t\sin^2\delta$, and $t_{xy} = \pm t\cos\delta\sin\delta$, 
with $t$ being the hopping parameter for the $\sigma$ bond. Here, the sign of $t_{xy}$ depends on the bond direction; hopping from A to B sublattice is positive (negative) if the bond points along $+x$ ($-x$). 
By taking the Fourier transform, we obtain Bloch Hamiltonian: 
\begin{align}
H_{k} = &\sum_{n=x,y,z}\left[2t_{nn}\cos \frac{k}{2} (\hat{1} - L_n^2)\sigma^0 \rho^x\right] + 2t_{xy}\sin \frac{k}{2} Q_{xy} \rho^y \nonumber \\
+ & (\lambda{\bm L} \cdot {\bm \sigma} - \Delta Q_{xy} - {\bm B}\cdot {\bm M})\rho^0
\label{eq:H_k}
.\end{align}
Here, ${\bm \rho}$ denotes Pauli matrix operators acting on the sublattice degrees of freedom, and $\hat{1}$ is the identity operator for the orbital sector. The Fourier transform is defined by the positions of atomic sites rather than the unit cells. 

\begin{figure}[t!]
    \centering
    \includegraphics[width=0.45\textwidth]{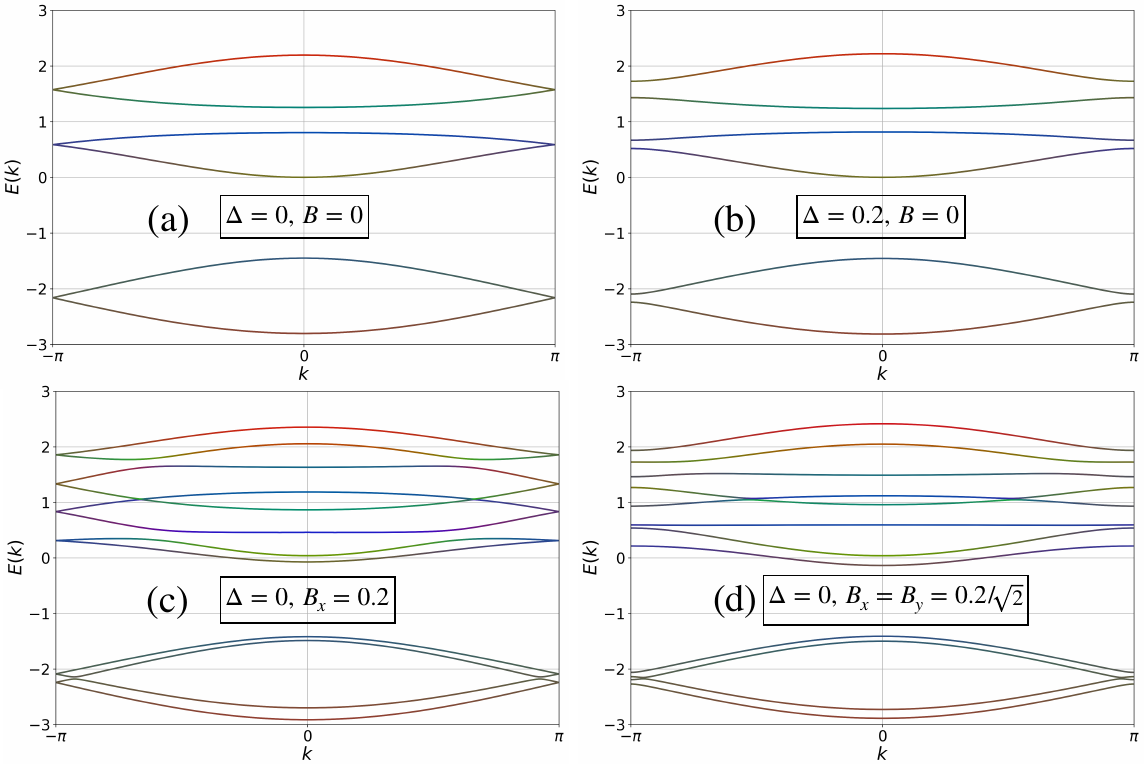}
    \caption{(Color online) Band structure of the zigzag chain model. Parameters are set to $t=-1$, $\delta = 30^\circ$, $\lambda = 1$, (a) $\Delta = B = 0$, (b) $\Delta = 0.2$, (c) $B_x = 0.2$, (d) $B_x = B_y = 0.2/\sqrt{2}$. Colors represent orbital weights: red, green, and blue for $p_x$, $p_y$, and $p_z$ orbitals.}
    \label{fig:Dispersion}
    \end{figure}

To examine gap opening, we compute the energy dispersion for $t = -1$, $\delta = 30^\circ$, and $\lambda = 1$. Figures~\ref{fig:Dispersion}(a)--(d) show the results. In Fig.~\ref{fig:Dispersion}(a), with $\Delta = B = 0$, the dispersion is degenerate at $k = \pi$. In Fig.~\ref{fig:Dispersion}(b), quadrupole order with $\Delta = 0.2$ opens a gap at $k = \pi$. The total angular momentum $j=1/2$ subspace near $E_k \simeq -2$ can be mapped onto the Su--Schrieffer--Heeger model~\cite{Su1979-ks}, featuring alternating hopping terms proportional to $\Delta t_{xy}/\lambda$. Other bands can also be mapped similarly if no band crossing occurs. 
In Fig.~\ref{fig:Dispersion}(c), applying a magnetic field in the $x$ axis with $B_x = 0.2$ lifts the Kramers degeneracy, but not the sublattice degeneracy at $k=\pi$. In Fig.~\ref{fig:Dispersion}(d), applying an in-plane magnetic field in the $xy$ plane with $B_x = B_y = 0.2/\sqrt{2}$ opens a gap at $k = \pi$. 

The bond-alternating order responsible for band gap opening is represented by a term $\propto \sin \frac{k}{2}\rho^y$, and its effective realization requires taking the trace over the $Q_{xy}=-(L_xL_y+L_yL_x)$ sector in the second term of Eq.~(\ref{eq:H_k}). Therefore, it is necessary, at least indirectly, to induce orbital quadrupole $Q_{xy}$ for opening a gap. For example, spin–orbit coupling is not required for the magnetic-field-induced gap opening, while it is necessary for ferromagnetic or ferroquadrupole orders of purely spin origin to induce a band gap.

We now discuss implications for real materials. In CeCoSi, Ce site positions represent an antiferroic arrangement of electric dipole $Q_z$ at ${\bm q}=(2\pi,0,0)$ and $(0,2\pi,0)$, and a ferroquadrupole order of the $(Q_{zx}, Q_{yz})$ type appears at a low temperature. \cite{Matsumura2022-ky}. This situation matches the zigzag chain model, with gaps opening at $k_x = \pi$ and $k_y = \pi$, consistent with the observed resistivity increase below the transition temperature~\cite{Lengyel2013-tn}. 
In ${\rm UCu_2Sn}$, U atoms form an ABAB-stacked triangular lattice, interpretable as electric octupole $Q_{y^3 - 3yx^2}$ order at ${\bm q} = (0,0,2\pi)$. A ferroic quadrupole order $Q_{x^2 - y^2}$~\cite{Suzuki2000-kb} couple to this and induces an electric dipole $Q_y({\bm q})$. Since the dipole is perpendicular to the wavevector, a gap may not open across an entire plane. Nevertheless, gaps can form at high-symmetry points, lines, or planes depending on orbital characters, which can contribute to increasing resistivity~\cite{Takabatake1998-ol}. 
Because ferroquadrupole order parameters can couple quadratically to magnetic fields, such systems may exhibit large anisotropic magnetoresistance near the transition temperature, suggesting fruitful directions for future experiments. 

In summary, we have shown that contrary to conventional expectations, zone-boundary gaps can be induced by ferroic orders and external magnetic fields. This has been explicitly demonstrated in a simple zigzag chain model. The results not only help resolve puzzles in existing experimental data but also offer prospects for magnetic-field control of transport phenomena.

{\it Acknowledgement.---}This work was supported by JSPS KAKENHI (Grant No. JP23K20824).


\begin{thebibliography}{7}%
\makeatletter
\providecommand \@ifxundefined [1]{%
 \@ifx{#1\undefined}
}%
\providecommand \@ifnum [1]{%
 \ifnum #1\expandafter \@firstoftwo
 \else \expandafter \@secondoftwo
 \fi
}%
\providecommand \@ifx [1]{%
 \ifx #1\expandafter \@firstoftwo
 \else \expandafter \@secondoftwo
 \fi
}%
\providecommand \natexlab [1]{#1}%
\providecommand \enquote  [1]{``#1''}%
\providecommand \bibnamefont  [1]{#1}%
\providecommand \bibfnamefont [1]{#1}%
\providecommand \citenamefont [1]{#1}%
\providecommand \href@noop [0]{\@secondoftwo}%
\providecommand \href [0]{\begingroup \@sanitize@url \@href}%
\providecommand \@href[1]{\@@startlink{#1}\@@href}%
\providecommand \@@href[1]{\endgroup#1\@@endlink}%
\providecommand \@sanitize@url [0]{\catcode `\\12\catcode `\$12\catcode
  `\&12\catcode `\#12\catcode `\^12\catcode `\_12\catcode `\%12\relax}%
\providecommand \@@startlink[1]{}%
\providecommand \@@endlink[0]{}%
\providecommand \url  [0]{\begingroup\@sanitize@url \@url }%
\providecommand \@url [1]{\endgroup\@href {#1}{\urlprefix }}%
\providecommand \urlprefix  [0]{URL }%
\providecommand \Eprint [0]{\href }%
\providecommand \doibase [0]{http://dx.doi.org/}%
\providecommand \selectlanguage [0]{\@gobble}%
\providecommand \bibinfo  [0]{\@secondoftwo}%
\providecommand \bibfield  [0]{\@secondoftwo}%
\providecommand \translation [1]{[#1]}%
\providecommand \BibitemOpen [0]{}%
\providecommand \bibitemStop [0]{}%
\providecommand \bibitemNoStop [0]{.\EOS\space}%
\providecommand \EOS [0]{\spacefactor3000\relax}%
\providecommand \BibitemShut  [1]{\csname bibitem#1\endcsname}%
\let\auto@bib@innerbib\@empty
\bibitem [{\citenamefont {Ashcroft}\ and\ \citenamefont
  {Mermin}(1976)}]{Ashcroft_Mermin}%
  \BibitemOpen
  \bibfield  {author} {\bibinfo {author} {\bibfnamefont {N.~W.}\ \bibnamefont
  {Ashcroft}}\ and\ \bibinfo {author} {\bibfnamefont {N.~D.}\ \bibnamefont
  {Mermin}},\ }\href@noop {} {\emph {\bibinfo {title} {Solid State Physics}}}\
  (\bibinfo  {publisher} {Holt, Rinehart and Winston},\ \bibinfo {address} {New
  York},\ \bibinfo {year} {1976})\BibitemShut {NoStop}%
\bibitem [{\citenamefont {Takabatake}\ \emph {et~al.}(1998)\citenamefont
  {Takabatake}, \citenamefont {Shirase}, \citenamefont {Katoh}, \citenamefont
  {Echizen}, \citenamefont {Sugiyama},\ and\ \citenamefont
  {Osakabe}}]{Takabatake1998-ol}%
  \BibitemOpen
  \bibfield  {author} {\bibinfo {author} {\bibfnamefont {T.}~\bibnamefont
  {Takabatake}}, \bibinfo {author} {\bibfnamefont {M.}~\bibnamefont {Shirase}},
  \bibinfo {author} {\bibfnamefont {K.}~\bibnamefont {Katoh}}, \bibinfo
  {author} {\bibfnamefont {Y.}~\bibnamefont {Echizen}}, \bibinfo {author}
  {\bibfnamefont {K.}~\bibnamefont {Sugiyama}}, \ and\ \bibinfo {author}
  {\bibfnamefont {T.}~\bibnamefont {Osakabe}},\ }\href {\doibase
  10.1016/S0304-8853(97)00347-8} {\bibfield  {journal} {\bibinfo  {journal} {J.
  Magn. Magn. Mater.}\ }\textbf {\bibinfo {volume} {177-181}},\ \bibinfo
  {pages} {53} (\bibinfo {year} {1998})}\BibitemShut {NoStop}%
\bibitem [{\citenamefont {Lengyel}\ \emph {et~al.}(2013)\citenamefont
  {Lengyel}, \citenamefont {Nicklas}, \citenamefont {Caroca-Canales},\ and\
  \citenamefont {Geibel}}]{Lengyel2013-tn}%
  \BibitemOpen
  \bibfield  {author} {\bibinfo {author} {\bibfnamefont {E.}~\bibnamefont
  {Lengyel}}, \bibinfo {author} {\bibfnamefont {M.}~\bibnamefont {Nicklas}},
  \bibinfo {author} {\bibfnamefont {N.}~\bibnamefont {Caroca-Canales}}, \ and\
  \bibinfo {author} {\bibfnamefont {C.}~\bibnamefont {Geibel}},\ }\href
  {\doibase 10.1103/PhysRevB.88.155137} {\bibfield  {journal} {\bibinfo
  {journal} {Phys. Rev. B}\ }\textbf {\bibinfo {volume} {88}},\ \bibinfo
  {pages} {155137} (\bibinfo {year} {2013})}\BibitemShut {NoStop}%
\bibitem [{\citenamefont {Suzuki}\ \emph {et~al.}(2000)\citenamefont {Suzuki},
  \citenamefont {Ishii}, \citenamefont {Okuda}, \citenamefont {Katoh},
  \citenamefont {Takabatake}, \citenamefont {Fujita},\ and\ \citenamefont
  {Tamaki}}]{Suzuki2000-kb}%
  \BibitemOpen
  \bibfield  {author} {\bibinfo {author} {\bibfnamefont {T.}~\bibnamefont
  {Suzuki}}, \bibinfo {author} {\bibfnamefont {I.}~\bibnamefont {Ishii}},
  \bibinfo {author} {\bibfnamefont {N.}~\bibnamefont {Okuda}}, \bibinfo
  {author} {\bibfnamefont {K.}~\bibnamefont {Katoh}}, \bibinfo {author}
  {\bibfnamefont {T.}~\bibnamefont {Takabatake}}, \bibinfo {author}
  {\bibfnamefont {T.}~\bibnamefont {Fujita}}, \ and\ \bibinfo {author}
  {\bibfnamefont {A.}~\bibnamefont {Tamaki}},\ }\href {\doibase
  10.1103/PhysRevB.62.49} {\bibfield  {journal} {\bibinfo  {journal} {Phys.
  Rev. B}\ }\textbf {\bibinfo {volume} {62}},\ \bibinfo {pages} {49} (\bibinfo
  {year} {2000})}\BibitemShut {NoStop}%
\bibitem [{\citenamefont {Matsumura}\ \emph {et~al.}(2022)\citenamefont
  {Matsumura}, \citenamefont {Kishida}, \citenamefont {Tsukagoshi},
  \citenamefont {Kawamura}, \citenamefont {Nakao},\ and\ \citenamefont
  {Tanida}}]{Matsumura2022-ky}%
  \BibitemOpen
  \bibfield  {author} {\bibinfo {author} {\bibfnamefont {T.}~\bibnamefont
  {Matsumura}}, \bibinfo {author} {\bibfnamefont {S.}~\bibnamefont {Kishida}},
  \bibinfo {author} {\bibfnamefont {M.}~\bibnamefont {Tsukagoshi}}, \bibinfo
  {author} {\bibfnamefont {Y.}~\bibnamefont {Kawamura}}, \bibinfo {author}
  {\bibfnamefont {H.}~\bibnamefont {Nakao}}, \ and\ \bibinfo {author}
  {\bibfnamefont {H.}~\bibnamefont {Tanida}},\ }\href {\doibase
  10.7566/JPSJ.91.064704} {\bibfield  {journal} {\bibinfo  {journal} {J. Phys.
  Soc. Jpn.}\ }\textbf {\bibinfo {volume} {91}},\ \bibinfo {pages} {064704}
  (\bibinfo {year} {2022})}\BibitemShut {NoStop}%
\bibitem [{\citenamefont {Su}\ \emph {et~al.}(1979)\citenamefont {Su},
  \citenamefont {Schrieffer},\ and\ \citenamefont {Heeger}}]{Su1979-ks}%
  \BibitemOpen
  \bibfield  {author} {\bibinfo {author} {\bibfnamefont {W.~P.}\ \bibnamefont
  {Su}}, \bibinfo {author} {\bibfnamefont {J.~R.}\ \bibnamefont {Schrieffer}},
  \ and\ \bibinfo {author} {\bibfnamefont {A.~J.}\ \bibnamefont {Heeger}},\
  }\href {\doibase 10.1103/PhysRevLett.42.1698} {\bibfield  {journal} {\bibinfo
   {journal} {Phys. Rev. Lett.}\ }\textbf {\bibinfo {volume} {42}},\ \bibinfo
  {pages} {1698} (\bibinfo {year} {1979})}\BibitemShut {NoStop}%
\bibitem [{\citenamefont {Hayami}\ and\ \citenamefont
  {Kusunose}(2024)}]{Hayami2024-vw}%
  \BibitemOpen
  \bibfield  {author} {\bibinfo {author} {\bibfnamefont {S.}~\bibnamefont
  {Hayami}}\ and\ \bibinfo {author} {\bibfnamefont {H.}~\bibnamefont
  {Kusunose}},\ }\href {\doibase 10.7566/JPSJ.93.072001} {\bibfield  {journal}
  {\bibinfo  {journal} {J. Phys. Soc. Jpn.}\ }\textbf {\bibinfo {volume}
  {93}},\ \bibinfo {pages} {072001} (\bibinfo {year} {2024})}\BibitemShut
  {NoStop}%
\end{thebibliography}
%

\end{document}